# MIPR:Automatic Annotation of Medical Images with Pixel Rearrangement

Pingping Dai[1] [*], Haiming Zhu[1] [*], Shuang Ge[1], Ruihan Zhang[1], Xiang Qian[1], Xi Li[2] [†], Kehong Yuan[1] [†]

[1]Tsinghua University   [2]Peking University Shenzhen Hospital

**Abstract.** Most of the state-of-the-art semantic segmentation reported in recent years is based on fully supervised deep learning in the medical domain. However, the high-quality annotated datasets require intense labor and domain knowledge, consuming enormous time and cost. Previous works that adopt semi-supervised and unsupervised learning are proposed to address the lack of annotated data through assisted training with unlabeled data and achieve good performance. Still, these methods can not directly get the image's annotation as doctors do. In this paper, inspired by self-training of semi-supervised learning, we propose a novel approach to solve the lack of annotated data from another angle, called medical image pixel rearrangement (short in MIPR). The MIPR combines *image-editing* and *pseudo-label* technology to obtain labeled data. As the number of iterations increases, the edited image is similar to the original image, and the labeled result is similar to the doctor's annotation. Therefore, the MIPR is to get labeled pairs of data directly from amounts of unlabeled data with pixel rearrangement, which is implemented with a designed conditional Generative Adversarial Networks and a segmentation network. Experiments on the ISIC18 show that the effect of the data annotated by our method for segmentation task is is equal to or even better than that of doctors' annotations.

## 1    Introduction

Semantic segmentation of medical images [1,2,3] is an essential and active task in medical imaging diagnosis. Many effective medical image segmentation methods based on deep learning have been proposed in recent years. e.g., segmentation frameworks [4,5,6], loss functions [7,8,32], and training tricks [9,10]. However, medical images still face notoriously data-hungry in deep learning. The main reasons are as follows, (a) Obtaining high-quality annotated datasets requires intense labor and domain knowledge, consuming enormous time and cost in the medical domain. (b) Differences in different imaging equipment and image processing methods may produce wide variations in resolution, image noise, and tissue appearance, which exacerbates the problem of limited labeled data.(c) It is difficult for images of rare diseases to form large public datasets as COCO [11], ADE20K [12].

To improve data-efficiency learning for semantic segmentation of medical images , many effective approaches have been proposed from different perspectives, such as using traditional data augmentation methods [13,14,15], utilizing unlabeled data for semi-supervised learning [20], generating new data by the generative model [21]. Particularly, Isensee F et al.[13] preprocessed data with traditional methods include crop, resize, resample, normalization, elastic deformation, etc. However, these functions are limited to labeled data and can be highly sensitive to the choice of parameters. Yang et al.[22] proposed TumorCP, which automatically 'copy' tumors and 'paste' them on random background online. The copy-paste data augmentation method is limited to the inherent anatomical structures and the context information around the lesions in medical images. Many previous works [23,24] have been studied transfer learning between CT and MRI, which need high-quality data. Wang X et al.[25] designed a semi-supervised self-training framework, utilizing unlabeled data and pseudo labeling method to segment images based on the low data regime. The performance of the semi-supervised learning and traditional data augmentation method is proportional to the amount of labeled data. Therefore, many works [21,27]focused on generating new training data. For example, Pérez A D et al.[27] generated synthetic eye fundus image from corresponding synthetic eye vessels. Foroozandeh M et al. [21] used two-stage GANs to generate pairs of images and masks for brain tumor segmentation. The high-quality synthesized images rely on large amounts of data to train the model well. Singh N K et al. [28] showed that a large section of research has focussed on the application of GANs in medical image synthesis of MRI imaging ,because the large number of MRI data set available in the public domain allowing researchers to have a surplus sample size for better model training. Moreover, Skandarani Y et al.[26] demonstrated that the medical information richness of synthesized images is lower than the clinical data's real images.

Our proposed approach, MIPR is to solve the problem of data-hungry from a new perspective. Inspired by semi-supervised self-training and conditional GANs, MIPR makes automatic annotation of medical images with pixel rearrangement, which only needs a few labeled data and a large of unlabeled data. In fact, it is common that the amount of unlabeled data is much more than the labeled data. Different from previous works of semi-supervised self-training for segmentation tasks[37,38], MIPR uses the 'pseudo label' to generate the 'pseudo image' by rearranging real image pixels according to the pseudo-label guide. With emphasis: 'pseudo label' is the correct annotation of 'pseudo image' in each iteration. Moreover, MIPR adds stronger constraints and new strategies for conditional GANs training to enable GANs performance well on the low data regime.

Our contributions are summarized as follows:
1. MIPR is a new framework to improve data-efficiency learning from a new perspective, which greatly reduces the expensive and time-consuming work for meidical data annotations.
2. MIPR retains the richness of the original medical image information, including distribution, diversity , and details.

3. We adopt a new GAN training strategy to avoid the problem of data over-fitting and details loss on the low data regime, which combines data augmentation and strong conditional constraints.

## 2 Method

### 2.1 Approach Overview

MIPR is a novel semi-supervised learning framework to automatically annotate medical images based on the low data regime. In this paper, we define the set of labeled data as $X_1$, the set of unlabeled data as $X_2$. Figure 1 shows the main steps: 1)Get manual annotation of a few images $X_1$; 2)Train the label-to-image model $G$ with $X_1$(top of (2)); 3)Train the semantic segmentation model $S$ with $X_1$(button of (2)); 4)Get the image's ($x_i$~ $X_2$) pseudo label $t_i$ with model $S$(left of (3)); 5)Get a pair of labeled data ($x'_i$, $t'_i$) with model $G$(right of (3)). When step1)~step2) are satisfied, step3)~step5) are defined as an iterative process of automatic annotation. At the same time, training data set to add labeled data ($x'_i$, $t'_i$).

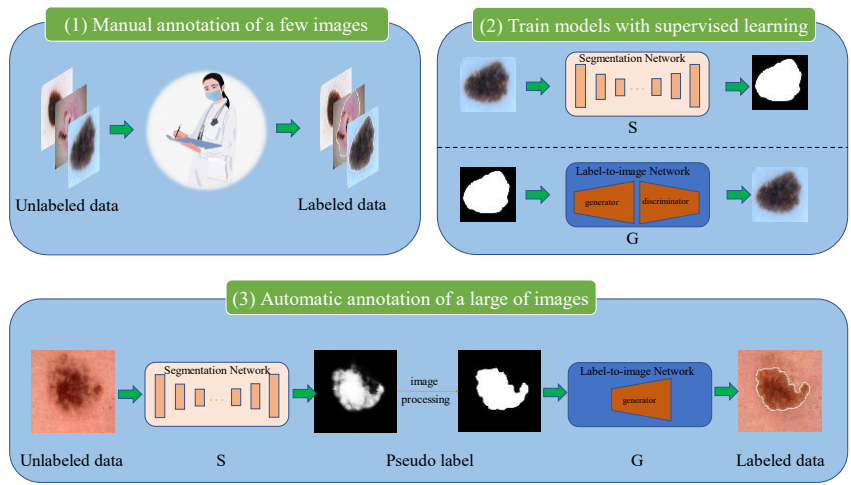

**Figure 1**: An overview of our proposed method. We use a few labeled data to train the semantic segmentation network $S$ and the label-to-image network $G$ with supervised learning. For a large of unlabeled medical images, 1) The pseudo label is obtained through network $S$. 2) Do image processing for the pseudo label; 3)A pair of image and label are obtained from the network $G$.

### 2.2 The segmentation network $S$

The segmentation network $S$ can be adopted in any semantic segmentation framework, such as FCN, Unet, Deeplabv3, etc. The main role of network $S$ is to generate pseudo labels and provide constraints on the shape and location of lesions for the image generated by network *G* . The training process uses a fully supervised approach, and The performance of MIPR can be improved by using data augmentation method and suitable network structure.

### 2.3 The label-to-image network *G*

We use the conditional GAN to generate images from labels. The architecture of the discriminator follows the one used in the GauGAN[29] , which uses a multi-scale design with the Spectral Norm. As Figure 2(a)shows, the generator is an image-to-image architecture such as the pix2pix model[30]，composed of an encoder and decoder.. The encoder consists of convolution blocks, and the decoder consists of SPADE+ blocks.

**The SPADE+ block.** Inspired by the SPADE[29], which can better preserve semantic information against common normalization layers, we add a new path to add details of generated images. Figure 2(b) shows, the SPADE+ block contains SPADE block and detailed path. The γ and β parameters In the SPADE block provide a spatially-variant affine transformation learned from the corresponding semantic mask to modulate the activation map. However, it is challenging to generate details in medical images only based on a simple semantic map. For example, in the dataset ISIC18, hair is also an essential component besides skin and lesions. We add a new path α based on the SPADE block to increase the details from the Sobel operator.

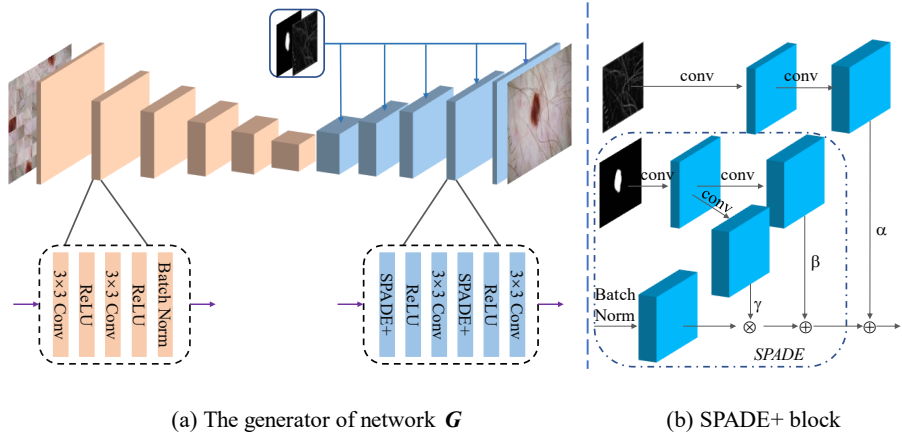

(a) The generator of network *G*   (b) SPADE+ block

**Figure 2**. (a)The generator's architecture of network G such as pix2pixHD[30]; (b)The SPADE+ block preserves semantic information and details of generated images.

**The pixel Rearrangement**. The deep feature maps of the encoder contain rich semantic information of the input image. Unlike the style-guide image synthesis in the SPADE work[29], our method directly generates synthesis images from the encoder's feature map rather than the reparametrization trick[31]. The advantage of our method is that it makes the distribution of the synthesized image more similar to the input and requires less labeled data and training cost. The biggest difference with image editing is that our goal is to recover input information from feature maps without loss and get its mask, so we call this process pixel rearrangement.

### 2.4 Train models with the data augmentation

It is a great challenge to train models well on a low data regime. especially for the Generative Adversarial Networks(GAN). Traditional data augmentation is the most common and effective solution, including rotation, flip, crop, resize, normalization, etc. Moreover, we adopt a simple and effective augmentation method in this paper, called patch-shuffle data augmentation. As Figure 3 shows, for example, when K=2, we can get six images of different local spatial structures, which is an effective data augmentation method. For training the network G, we adopt different values of N between the encoder's input and label. It performs better when the patch shuffle is combined with rotation, color jitter, etc.

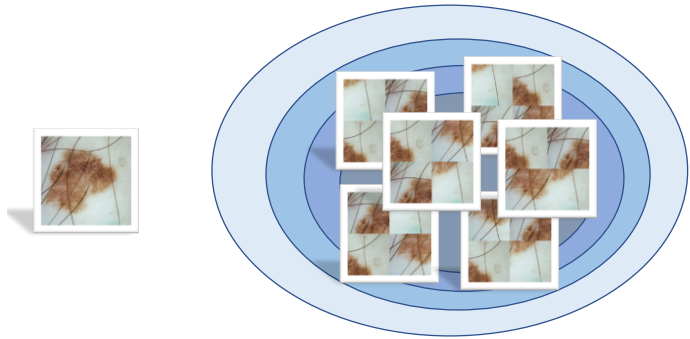

**Figure 3**. When K=2, we will get 6 images with different structures by the patch-shuffle data augmentation method.

## 3 Experiments and Discussion

### 3.1 Experiment settings

We evaluate MIPR on ISIC18[33], a publicly available skin lesion segmentation dataset was published by the International Skin Imaging Collaboration (ISIC). The ISIC18 contains 2594 images, and the size is ranged from 720×540 to 6708×4439. To eliminate the similarity between the adjacent images of the original name, we rename the images randomly and resize all images as 256×256. We split the dataset into three parts: 1)200 images as the labeled data for training; 2)1800 images as the unlabeled data (We assume that these images are not manually annotated); 3) 594 images as the labeled data for test.

All experiments are conducted on an NVIDIA with 8 12 GB TITAN 2080Ti GPU. The training hyperparameters of network $S$ : epoch is 100, initial lr is 0.001, and has been changed rate by cosine annealing. For the network $G$, we use the training hyperparameters of publicly available SPADE[29] codebase for implementation but 500 epochs.

Evaluation metrics of segmentation accuracy are based on: 1) Dice similarity coefficient (DSC) measures the overlap of model's prediction and ground, formulated as DSC=(2TP)/(FP+2TP+FN). 2) Accuracy measures the proportion of correct predictions among the total number of cases examined, formulated as ACC=(TP+TN) /(TP+FP+TN+FN). where TP = True positive; FP = False positive; TN = True negative; FN = False negative.

### 3.2 Image Quality

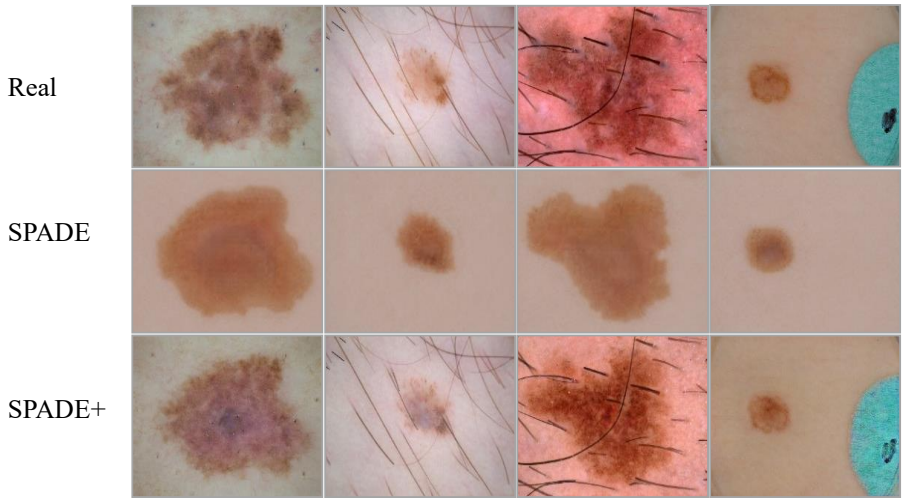

**Figure 4**. Visual comparisons of SPADE+(**ours**) and SPADE with real images based on the *low data regime* on ISIC18.

As mentioned before, it is a great challenge to train GANs well on a low data regime. One reason is that models are easy to overfit, making it challenging to generate images as diverse as clinical images. The other reason is that it is challenging to generate details. Although recent works [34,35] generate details, it is based on large amounts of data. Figure 4 visually shows the images generated by a GAN are made of SPADE block or SPADE+ block. The images in the third row(**ours**) have much more details than those in the second row(SPADE). In other words, our method can generate images with details closer to the real image based on the low data regime.

To further analyze the quality of the generated images, we used t-SNE to describe the distribution of data, with the feature vectors coming from the VGG19's (pre-trained on ImageNet) last fully connected layer. As shown in Figure 5, The images distribution generated by our method is closer to real clinical data.

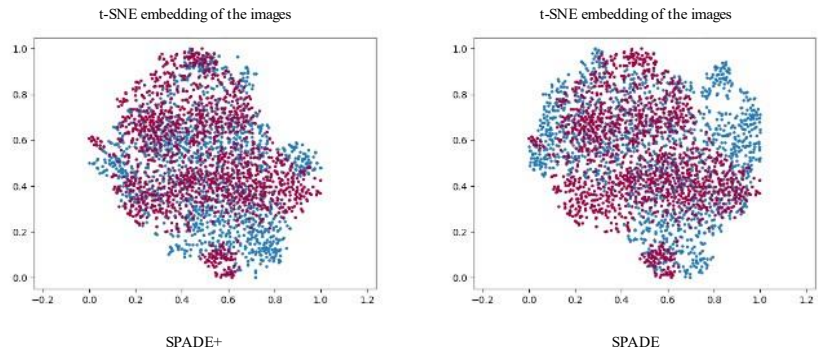

SPADE+        SPADE

**Figure 5**. Comparison of t-SNE projection of images generated by a GAN is made of SPADE block or SPADE+ block(**ours**), with a VGG19 trained on ImageNet. Red points: Real images, Blue points: Generated images.

### 3.3  Evaluate MIPR on the segmentation task

**Table 1**. Evaluate MIPR on the segmentation task, 'Doctor' represents manual annotations from the publicly available ISIC18, 'data amounts' means the number of training datasets, 'iter' means the number of iterations of MIPR.

| Model | Method[iter] (data amounts) | DSC±std | Acc±std |
|---|---|---|---|
| Unet | Doctor    (1800) | **68.41±1.71** | **90.46±1.01** |
|  | MIPR[1] (1513) | 66.52±2.56 | 89.78±1.17 |
|  | MIPR[2] (1691) | 67.61±1.64 | 90.29±0.90 |
|  | MIPR[3] (1781) | 68.17±1.85 | 89.82±1.04 |
| Att-Unet | Doctor    (1800) | 69.94±1.67 | **91.37±1.04** |
|  | MIPR[1] (1622) | 68.57±2.39 | 90.81±1.21 |
|  | MIPR[2] (1780) | 68.70±2.52 | 90.54±1.10 |
|  | MIPR[3] (1800) | **70.09±1.91** | 90.93±1.06 |

We use Unet [4] and Att-Unet[36] as the network $S$ to evaluate the effectiveness of the MIPR in different iterations. Compared with the data manually annotated, the DSC score and ACC score of the data obtained by MIPR are close to them. It is surprising that MIPR even performances better than manual annotation in the case of *Att-Unet/MIPR[3]*.

Our analysis shows that manual annotation also has a subjective error.

### 3.4 Visualization

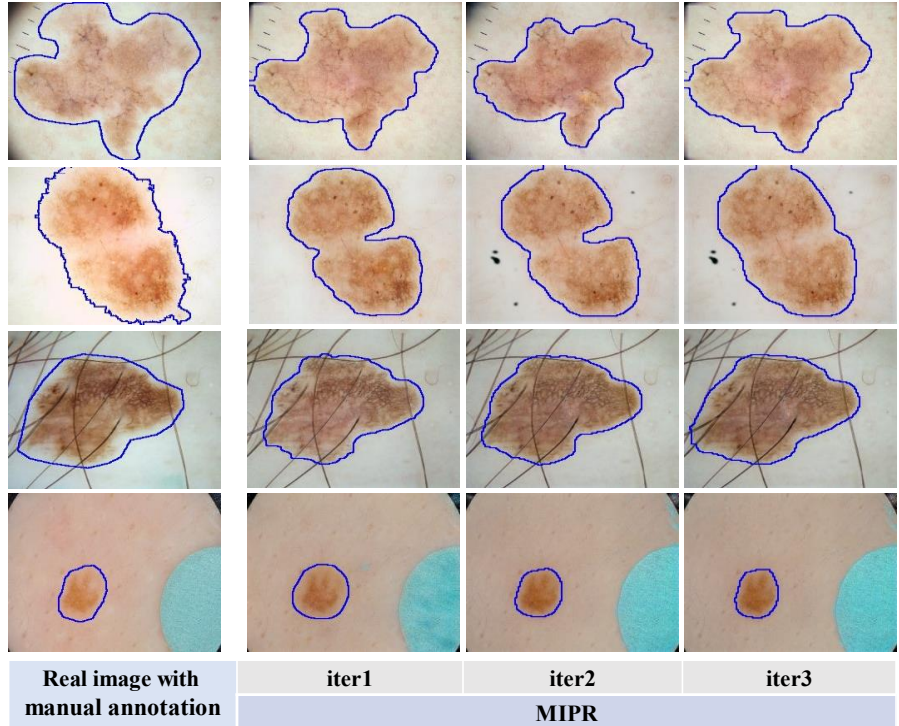

**Figure 6**. Automatic annotation results on ISIC18, MIPR successfully automated annotated data, similar to manual annotation results.

Figure 6 shows the automatic annotation results with MIPR in different iterations. We get similar results with manual annotation and highly retain the details of the original image.

## 4 Conclusion and Future Works

The contribution of this paper is to propose a method from a new perspective, MIPR, which aims to solve the lack of annotation data in medical domain. MIPR is designed with a semi-supervised self-training framework and a GAN network. Experiments proved that MIPR saves the cost and time of data annotation, and at the same time ensures the quality of annotation and the richness of original data information. For the future works, the MIPR framework is not limited to annotate medical images. Changing constraints can generate data and improve the application of deep learning in medical image processing from the perspective of data.

# References

1. Menze B H, Jakab A, Bauer S, et al. The multimodal brain tumor image segmentation benchmark (BRATS)[J]. IEEE transactions on medical imaging, 34(10): 1993-2024( 2014).
2. Heller N, Isensee F, Maier-Hein K H, et al. The state of the art in kidney and kidney tumor segmentation in contrast-enhanced CT imaging: Results of the KiTS19 challenge[J]. Medical image analysis, 67: 101821(2021).
3. Jha D, Smedsrud P H, Riegler M A, et al. Kvasir-seg: A segmented polyp dataset[C]. International Conference on Multimedia Modeling. Springer, Cham, pp. 451-462( 2020).
4. Ronneberger O, Fischer P, Brox T. U-net: Convolutional networks for biomedical image segmentation[C]. International Conference on Medical image computing and computer-assisted intervention. Springer, Cham, pp. 234-241(2015).
5. Zhou Z, Rahman Siddiquee M M, Tajbakhsh N, et al. Unet++: A nested u-net architecture for medical image segmentation[M]. Deep learning in medical image analysis and multimodal learning for clinical decision support. Springer, Cham, pp. 3-11(2018).
6. Chen J, Lu Y, Yu Q, et al. Transunet: Transformers make strong encoders for medical image segmentation[J]. arXiv preprint arXiv:2102.04306(2021).
7. Lin T Y, Goyal P, Girshick R, et al. Focal loss for dense object detection[C]. Proceedings of the IEEE international conference on computer vision. pp. 2980-2988(2017).
8. Taghanaki S A, Zheng Y, Zhou S K, et al. Combo loss: Handling input and output imbalance in multi-organ segmentation. In CMIG, pp. 24-33(2019).
9. Zhao Y X, Zhang Y M, Liu C L. Bag of tricks for 3D MRI brain tumor segmentation[C]. International MICCAI Brainlesion Workshop. Springer, Cham,pp. 210-220(2019).
10. Koch T L, Perslev M, Igel C, et al. Accurate segmentation of dental panoramic radiographs with U-Nets[C]. 2019 IEEE 16th International Symposium on Biomedical Imaging. IEEE, pp. 15-19(2019).
11. Lin T Y, Maire M, Belongie S, et al. Microsoft coco: Common objects in context[C]. European conference on computer vision. Springer, Cham, pp. 740-755(2014).
12. Zhou B, Zhao H, Puig X, et al. Semantic understanding of scenes through the ade20k dataset[J]. International Journal of Computer Vision, pp. 302-321(2019).
13. Isensee F, Jäger P F, Kohl S A A, et al. Automated design of deep learning methods for biomedical image segmentation[J]. arXiv preprint arXiv:1904.08128(2019).
14. Sirinukunwattana K, Pluim J P W, Chen H, et al. Gland segmentation in colon histology images: The glas challenge contest[J]. Medical image analysis,pp.: 489-502(2017).
15. Dong H, Yang G, Liu F, et al. Automatic brain tumor detection and segmentation using U-Net based fully convolutional networks[C]. annual conference on medical image understanding and analysis. Springer, Cham, pp. 506-517(2017).
16. Zhao A, Balakrishnan G, Durand F, et al. Data augmentation using learned transformations for one-shot medical image segmentation[C]. Proceedings of the ieee/cvf conference on computer vision and pattern recognition. pp. 8543-8553(2019).
17. Hu X, Zeng D, Xu X, et al. Semi-supervised contrastive learning for label-efficient medical image segmentation[C]. International Conference on Medical Image Computing and Computer-Assisted Intervention. Springer, Cham,pp. 481-490(2021).
18. Zhou H Y, Yu S, Bian C, et al. Comparing to learn: Surpassing imagenet pretraining on radiographs by comparing image representations[C]. International Conference on Medical Image Computing and Computer-Assisted Intervention. Springer, Cham, pp. 398-407(202).
19. Chang Q, Qu H, Zhang Y, et al. Synthetic learning: Learn from distributed asynchronized discriminator gan without sharing medical image data[C]. Proceedings of the IEEE/CVF Conference on Computer Vision and Pattern Recognition. pp. 13856-13866(2020).

20. Liao X, Li W, Xu Q, et al. Iteratively-refined interactive 3D medical image segmentation with multi-agent reinforcement learning[C]. Proceedings of the IEEE/CVF conference on computer vision and pattern recognition. pp. 9394-9402(2020).
21. Foroozandeh M, Eklund A. Synthesizing brain tumor images and annotations by combining progressive growing GAN and SPADE. arXiv preprint arXiv:2009.05946(2020).
22. Yang J, Zhang Y, Liang Y, et al. TumorCP: A Simple but Effective Object-Level Data Augmentation for Tumor Segmentation[C]. International Conference on Medical Image Computing and Computer-Assisted Intervention. Springer, Cham, pp. 579-588(2021).
23. Jiang J, Hu Y C, Tyagi N, et al. Tumor-aware, adversarial domain adaptation from ct to mri for lung cancer segmentation[C]. International Conference on Medical Image Computing and Computer-Assisted Intervention. Springer, Cham, pp. 777-785(2018).
24. Wolterink J M, Dinkla A M, Savenije M H F, et al. Deep MR to CT synthesis using unpaired data[C]. International workshop on simulation and synthesis in medical imaging. Springer, Cham, pp. 14-23(2017).
25. Wang X, Chen H, Xiang H, et al. Deep virtual adversarial self-training with consistency regularization for semi-supervised medical image classification[J]. Medical image analysis, pp. 102010(2017).
26. Skandarani Y, Jodoin P M, Lalande A. Gans for medical image synthesis: An empirical study[J]. arXiv preprint arXiv:2105.05318( 2021).
27. Pérez A D, Perdomo O, Rios H, et al. A conditional generative adversarial network-based method for eye fundus image quality enhancement[C]. International Workshop on Ophthalmic Medical Image Analysis. Springer, Cham, pp. 185-194(2020).
28. Singh N K, Raza K. Medical image generation using generative adversarial networks: a review[J]. Health Informatics: A Computational Perspective in Healthcare, pp. 77-96(2017).
29. Park T, Liu M Y, Wang T C, et al. Semantic image synthesis with spatially-adaptive normalization[C]. Proceedings of the IEEE/CVF conference on computer vision and pattern recognition. pp. 2337-2346(2019).
30. Wang T C, Liu M Y, Zhu J Y, et al. High-resolution image synthesis and semantic manipulation with conditional gans[C]. Proceedings of the IEEE conference on computer vision and pattern recognition. pp. 8798-8807(2018).
31. D. P. Kingma and M. Welling. Auto-encoding variational bayes. In International Conference on Learning Representations (ICLR)(2014).
32. Milletari F, Navab N, Ahmadi S A. V-net: Fully convolutional neural networks for volumetric medical image segmentation[C]. 2016 fourth international conference on 3D vision (3DV). IEEE, pp. 565-571(2016).
33. Wang Z, Bovik A C, Sheikh H R, et al. Image quality assessment: from error visibility to structural similarity[J]. IEEE transactions on image processing, pp. 600-612(2014).
34. Codella N , Gutman D , Celebi M E , et al. Skin Lesion Analysis Toward Melanoma Detection: A Challenge at the 2017 International Symposium on Biomedical Imaging (ISBI), Hosted by the International Skin Imaging Collaboration(2017).
35. Karras T, Aittala M, Laine S, et al. Alias-free generative adversarial networks[J]. Advances in Neural Information Processing Systems(2021).
36. Oktay O, Schlemper J, Folgoc L L, et al. Attention u-net: Learning where to look for the pancreas[J]. arXiv preprint arXiv:1804.03999(2018).
37. Ke R, Aviles-Rivero A I, Pandey S, et al. A Three-Stage Self-Training Framework for Semi-Supervised Semantic Segmentation[J]. IEEE Transactions on Image Processing(2022).
38. Zhu Y, Zhang Z, Wu C, et al. Improving semantic segmentation via self-training[J]. arXiv preprint arXiv:2004.14960(2020).